\documentclass[
aps,%
12pt,%
final,%
notitlepage,%
oneside,%
onecolumn,%
nobibnotes,%
nofootinbib,%
superscriptaddress,%
noshowpacs,%
centertags]%
{revtex4}
\usepackage{mathtools}
\begin{document}
\selectlanguage{english}
\title{Ultralight Axion or Axion-Like Particle Dark Matter and 21-cm Aborption Signals in New Physics}

\author{\firstname{C.~R.}~\surname{Das}}

\email{das@theor.jinr.ru}

\affiliation{The Bogoliubov Laboratory of Theoretical Physics, International Intergovernmental Scientific Research Organization, Joint Institute for Nuclear Research, Dubna, Russia}


\begin{abstract}
{\bf Abstract -} A hypothetical particle known as the axion holds the potential to resolve both the cosmic dark matter riddle and particle physics' long-standing, strong CP dilemma. An unusually strong 21-cm absorption feature associated with the initial star formation era, i.e., the dark ages, may be due to ultralight axion dark matter ($\sim$10$^{-22}$ eV) at this time. The radio wave observation's 21-cm absorption signal can be explained as either anomalous baryon cooling or anomalous cosmic microwave background photon heating. Shortly after the axions or axion-like particles (ALPs) thermalize among themselves and form a Bose--Einstein condensate, the cold dark matter ALPs make thermal contact with baryons, cooling them. ALPs are thought to be the source of some new evidence for dark matter, as the baryon temperature at cosmic dawn was lower than predicted based on presumptions. The detection of baryon acoustic oscillations is found to be consistent with baryon cooling by dark matter ALPs. Simultaneously, under the influence of the primordial black hole and/or intergalactic magnetic fields, the dark radiation composed of ALPs can resonantly transform into photons, significantly heating up the radiation in the frequency range relevant to the 21-cm tests. When examining the 21-cm cosmology at redshifts $z$ between 200 and 20, we see that, when taking into account both heating and cooling options at the same time, heating eliminated the theoretical excess number of neutrino species, $\Delta N_{\rm eff}$, from the cooling effect.
\end{abstract}

\maketitle

\section{Ultralight axion-like dark matter}

Numerous experimental validations of dark matter's existence exist; nonetheless, a definitive comprehension of its nature remains elusive \cite{2017IJMPD..2630012F}. In light of the lack of substantial findings in the quest for weakly interacting massive particles (WIMPs) \cite{2017PhRvL.118b1303A,2018PhRvL.121k1302A,2021PhRvL.127z1802M}, there has been a notable shift in focus towards ultralight dark matter (DM) candidates, such as dark photons, quantum chromodynamic (QCD) axions, and other pseudo-Nambu--Goldstone bosons, postulated in some extensions of the standard model of particle physics, called axion-like particles (ALPs) \cite{2024arXiv240409036R}. It appears that the first galaxies were larger, brighter, and more numerous than predicted by WIMP-based models \cite{Stacy_McGaugh_Updated_WIMP_Exclusion_Diagram}. The QCD axion is a pseudo-scalar, originally proposed to tackle the strong CP problem as a theoretical particle \cite{1977PhRvL..38.1440P,1977PhRvD..16.1791P,1978PhRvL..40..223W,1978PhRvL..40..279W}, where ``CP'' denotes the interplay of charge conjugation symmetry and parity symmetry, has demonstrated potential as a viable candidate for dark matter \cite{1983PhRvL..50..925I}. A plausible hypothesis involves ultralight dark matter, as references to indicate that ALPs may serve as a potential candidate for dark matter \cite{MARSH20161,PhysRevD.95.043541,PRESKILL1983127,ABBOTT1983133,DINE1983137}. The mass is observationally constrained to $m_a < 10^{-3}$ eV \cite{Raffelt2008}, while other theories predict $m_a \sim 10^{-10}$ eV \cite{RevModPhys.82.557,PhysRevD.83.044026,PhysRevD.95.043541}. ALPs that appear in string-theory models (in the mass range $10^{-33}$ to $10^{-10}$ eV) \cite{PhysRevD.81.123530,MarkGoodsell2009}, for example, and are similarly connected to SM particles are also interesting candidates for DM \cite{2006JHEP...06..051S}. The early universe's topological object decay \cite{1982PhRvL..48.1867V,1982PhRvL..48.1156S} or the misalignment mechanism can produce axions or ALPs \cite{PRESKILL1983127,ABBOTT1983133,DINE1983137}.

The parameter ranges $10^{-33}$ eV $\leq m_a \leq 10^{-2}$ eV are particularly relevant for DM abundances \cite{MARSH20161,Rogers_2023}. ALPs with a mass of $10^{-27}$ eV exhibit early dark energy (DE) behavior, whereas those with a mass of $10^{-33}$ eV have typical DE behavior and can currently drive cosmic acceleration, and heavier ALPs behave as DM \cite{Grin2019Gravitational}. In this research, we suppose that ALPs make up ultralight dark matter (ULDM) \cite{2023NatAs...7..736A} mass around $\sim$10$^{-22}$ eV. This is driven by the ALP mass's absence of perturbative renormalization, which shields the tiny mass from radiative corrections. Because ALPs couple with photons, this kind of dark matter may be discovered. ALPs can be distinguished from heavier dark matter candidates by their small-scale signatures, which include solitons, interference patterns, and suppression of structure. Due to their gravitational imprint, ALPs in the currently permitted parameter space provide a variety of observational tests to aim for in the upcoming decade, changing conventional predictions for the black hole mass spectrum, pulsar timing, Lyman-$\alpha$ absorption by neutral hydrogen along quasar sightlines, galaxy clustering, and microwave background anisotropies.

The couplings between photons or ALPs and SM particles provide crucial instruments for searching for these lightweight particles; several sorts of studies explore photon-associated signals \cite{1983PhRvL..51.1415S,1985PhRvD..32.2988S,1982JETP...56..502O,1987PhRvL..59..759V,2015PhLB..747..331A,2013PhRvL.111d1302A}. References contain a thorough list of experimental restrictions for photons and ALPs \cite{2021PhRvD.104i5029C,ciaran_o_hare_2020_3932430}. Another useful method for searching for ALPs or dark photons is to investigate anomalous signals in a variety of astrophysical environments, such as the Sun, red giants and horizontal branch stars, neutron stars, white dwarfs, supernovae, quasars and blazars, globular clusters, gamma-ray constraints on ALP DM, and the cosmic microwave background (CMB) spectral distortion constraints on dark photons (\cite{2024NatCo..15..915A} and refs.~inside). The standard cosmology model accurately predicts the frequency and luminance temperature of the trough's minimum in relation to the radio background. This is due to the fact that the gas is cooled relative to the CMB solely through adiabatic expansion. Any deviation from this prediction is indicative of the existence of new physics. A tiny conversion rate can lead the CMB brightness temperature to double in the Rayleigh--Jeans tail frequency band, where the number of hidden photons is likely larger than the CMB photon number density.

\section{21-cm absorption signals and new physics}

The hydrogen atom spin-flip transition, also known as the 21-cm line or the ``neutral Hydrogen'' line, is significant in Big Bang cosmology as it provides the sole method to investigate the cosmological ``dark ages'' from recombination, when stable hydrogen atoms were initially created, to reionization. Considering the redshift, this line will be detected at frequencies ranging from 200 MHz to approximately 15 MHz on Earth. There are two possible uses for it. First, it can theoretically give a very accurate image of the matter power spectrum in the post-recombination phase by mapping the intensity of redshifted 21-cm light. Second, because neutral hydrogen ionized by star or quasar radiation will show up as holes in the 21-cm background, it can give an image of how the universe was re-ionized.

The radio wave observation's 21-cm absorption signal can be attributed to either anomalous baryon cooling or anomalous CMB photon heating. The cold dark matter (CDM) ALPs come into thermal contact with baryons, cooling them, and then thermalize among themselves, forming a Bose--Einstein condensate (BEC) \cite{PhysRevD.85.063520}. The EDGES collaboration's measurement \cite{2018Natur.555...67B,2019PhRvD.100l3005B} of a baryon temperature at cosmic dawn that is lower than expected under ``standard'' assumptions is regarded as further evidence that dark matter is made up, at least in part, of ALPs. Dark matter ALPs are discovered to chill baryons, which is compatible with observations of baryon acoustic oscillations \cite{SIKIVIE2019100289}. It is very susceptible to further heating or cooling, though. A model with additional cooling results in a deeper depth of absorption line. In particular, theories where dark matter decays or self-annihilates are heated, whereas models where dark matter particles interact with baryonic matter through the weak Coulomb-like force are cooled. Heating baryonic materials causes the redshifted 21-cm absorption line to become shallow or turn into an emission line. The Cosmic Dawn trough is caused by a combination of sophisticated physics and astrophysics. The Dark Ages, on the other hand, are considerably easier because astrophysics did not exist at the time. The cooling models effectively segregate lower frequencies and limit deviations from classic $\Lambda$CDM cosmology generated by new physics, including unanticipated interactions with dark matter. Observations of the redshifted 21-cm line during the Dark Ages have the potential to reveal more new physics in an unseen Farside period. This encompasses the decay and/or annihilation of dark matter, primordial black holes, cosmic strings, and early dark energy. 

\section{Baryon cooling and transformation into photons of ALP-like particles (ALPs)}

\subsection{Baryon Cooling}

ALP-induced cooling occurs because coherence amplifies scattering energy losses, resulting in hydrogen/baryon cooling. The total cooling amplitude is determined by the sum of scattering targets (also known as `individual' ALPs) that interfere constructively when coherent. Resulting cooling rate:
\begin{equation}
\Gamma_{\rm baryon\; cooling} \simeq 4\pi G m_an_al_a {E_H\over \Delta p} ,\label{eqn1}
\end{equation}
where $m_a$, $n_a$, and $l_a$ are ALP mass, number density, and correlation length, respectively. The energy and momentum dispersion terms for hydrogen are $E_H$ and $\Delta p$. Their correlation length is the $l_a\sim t_1(a(t)/a(t_1))$. During the QCD phase transition, the ALP mass effectively turns on at $t_1 \sim 2\times 10^{-7}$ s, and the scale factor is $a(t)$. When the particles' energy dispersion is less than their rate of relaxation, $\Gamma_a \simeq 4\pi G m_a^2n_al^2_a$ is applicable, using $\Delta p = l_a^{-1}$ and $E_H = m_a$ as suitable for the cold ALPs. It has been suggested that gravitational interactions allow this state to rethermalize over time while maintaining large-scale coherence. Also, relativistic axion states and photons have a relaxation rate of $\Gamma_{a/\gamma} \simeq 4\pi Gm_an_al_a$ due to gravitational interactions with cold ALPs at $E_H\sim \Delta p$. To differentiate it from the more frequently seen ``particle kinetic regime,'' which is defined by the requirement that the energy dispersion of the particles is considerable in relation to the relaxation rate, is the ``condensed regime.'' After $t_1$, the condensed regime contains cold dark matter ALPs. The cooling process starts once $\Gamma_{\rm cooling}(t)/H(t)\sim 1$, when redshift $200 > z > 20$. At $z=200$, CMB/hydrogen decoupling occurs, and at $z=20$, EDGES measurements begin. This defines the needed ALP-CDM density as a function of redshift. The Hubble rate, $H(t)$, is of order $5\times 10^{-7}$ at $t_1$. It grows with time as $a(t)^{-1}t$ and reaches one at roughly 500 eV photon temperature. At that point, the ALPs thermalize and form a BEC. Almost all ALPs transitions to the lowest energy state are possible. The correlation length increases and becomes of order of the horizon. In the linear regime of density perturbation evolution and within the horizon, the lowest energy state is time-independent, and no rethermalization is required for the ALPs to remain in that condition. In that circumstance, ALP BEC and conventional CDM are indistinguishable at all levels of observational interest. However, beyond first-order perturbation theory and/or upon entering the horizon, the ALPs rethermalize in an attempt to maintain the lowest energy state possible. ALP-BEC behaves differently from CDM, and the differences are noticeable. This defines $m_a$ as a function of redshift, or, equivalently, when hydrogen cooling begins \cite{SIKIVIE2019100289}. Since gravitational effects are hard to include in QFT, the same effect is unquestionably achievable in more generic ALP models through ALP self-interactions \cite{PhysRevD.92.103513}.

\subsection{ALPs to Photon Conversion and CMB Photon Heating}

The ALPs to photon conversion rate in the magnetic field:
\begin{widetext}
\begin{equation}
\Gamma_{a\; \to\; \gamma\; (\rm CMB\; heating)} \simeq -\ln \left[1-\left({\pi r g_a^2 B^2_\perp E_a \over m_a^2}\right)\right]\times\left.{4\sqrt{E_a r}\over m_a}\; \right|_{\; z\; =\; z_{\rm res}},
\end{equation}
\end{widetext}
where $r$ is the inverse expansion rate of the Hubble parameter $H$, $g_a$ represents the strength of the ALP--photon coupling, and $B_\perp$ denotes the strength of the magnetic field perpendicular to the momentum direction of ALP, with energy $E_a$ that passes through the region with a non-vanishing magnetic field. The plasma frequency $\omega_p(z)$ changes with Hubble expansion, and there is an epoch, denoted by redshift $z = z_{\rm res}$, where $\omega^2_p$ becomes equal to $m^2_a$ for a small ALP mass across $10^{-23}$--$10^{-22}$ eV. Such low-energy photons may encounter bremsstrahlung absorption above the redshift $z \gtrsim 2000$, constraining the resonant redshift to $z_{\rm res} \lesssim 2000$. In this scenario, ALP interacts with photons and can convert to photons under the magnetic fields of intergalactic and PBHs, dramatically increasing the radio band brightness temperature. As long as the resonant conversion happens before the redshifts $z \gtrsim 20$, the intensity of the observed 21-cm signal can be increased in comparison to the purely astrophysical effects. Here, ALPs' oscillation length is shorter than both the magnetic field's usual coherent length and the photon's mean free path \cite{2018PhLB..783..301M}. The ALPs decay generates additional radiation backgrounds, raising the baryon temperature.

\subsection{The Thermal Equilibrium}

The thermal equilibrium rate between baryon cooling and CMB photon heating is:
\begin{widetext}
\begin{equation}
\left|\; \Gamma_{\rm baryon\; cooling} \xrightleftharpoons[\circlearrowright]{\circlearrowleft} \Gamma_{a\; \to\; \gamma\; (\rm CMB\; heating)}\; \right|_{\;{\rm thermal\; equilibrium\;} (z\; =\; 200\; \Longleftrightarrow\; 20)} .\label{eqn3}
\end{equation}
\end{widetext}
For resonant conversion to occur before and around matter-radiation equality, the ALP mass should be in the range of $m_a \sim 10^{-22}$ eV. The resonant conversion of the ALP into photons occurs, as do the simultaneous cooling and heating processes, as well as the natural thermalization process according to the redshift density profile; this requires further studies with simulations. The chance of resonant conversion increases with redshift $z_{\rm res} \lesssim 1000$ due to the recombination effect, which causes a sharp jump around it. However, the thermal equilibrium rate of baryon cooling and heating occurs at about $z = 200$, when the gas decouples from the CMB radiation and begins adiabatically cooling until the first things form and heat up the gas at redshifts less than 30. At $z \approx 100$, the Hubble expansion increases the effectiveness of collisional coupling with gas and decreases, resulting in thermal equilibrium rate variations. Again, this is determined by the size and strength of the primordial PBH's and/or intergalactic magnetic field.

\section{The effective number of neutrino species}

The majority of the ALPs remain in the ground state even after they are heated to the same temperature as the photons. In the ground state, the ALPs exhibit cold dark matter behavior. One bosonic degree of freedom is added to radiation by the ALPs in the excited states. The effective number $N_{\rm eff}$ of thermally excited neutrino degrees of freedom is frequently used to express the universe's radiation content. This number is defined by:
\begin{equation}
\rho_{\rm rad} = \rho_\gamma\left[1+N_{\rm eff}{7\over 8}\left({4\over 11}\right)^{4\over 3}\right].\label{eqn4}
\end{equation}
$N_{\rm eff} = 3.044$ is predicted by the standard cosmological model with conventional cold dark matter. This value is marginally greater than 3 due to the minor heating of the three neutrinos during $e^+e^-$ annihilation. According to the suggested scenario, the three regular neutrinos' contribution is increased since the photons have been cooled in comparison to them, in addition to the additional species of radiation from thermally excited ALPs. So, if we consider $\rho_{\rm rad} = \rho_\gamma + \rho_a + \rho_\nu$, it yields twice the predicted $N_{\rm eff}$ value around 7, with cosmological parameters within their standard values.
\begin{eqnarray}
  \rho_{\rm rad} &=& \rho_\gamma + \rho_a + \rho_\nu\nonumber\\
  &=& \rho_\gamma\left[1+{1\over 2}+3.044 {7\over 8}\left({4\over 11}\right)^{4\over 3}{3\over 2}\right] ,\label{eqn5}
\end{eqnarray}
where $\rho_\gamma$, $\rho_a$, and $\rho_\nu$ are the energy density of photons, ALPs, and neutrinos, respectively, and $\rho_{\rm rad}$ is the energy density in radiation overall \cite{PhysRevD.85.063520}.

\subsection{Thermal Equilibrium and $\Delta N_{\rm eff}$}

The heating impact of ALP to photon conversion can eliminate this aberration. From \cite{2018PhLB..783..301M}, the energy fraction of the ALP radiation in the EDGES frequency value is $f^{\rm (EDGES)}_a=0.4$. Because of the ALP--photon conversion, we must raise the photon energy density in the EDGES range by $\cal{O}$(1). So, the conversion probability of ALP to photon:
\begin{equation}
P_{a\; \to\; \gamma}\left(E_a\right)\sim2\times\left(f^{\rm (EDGES)}_\gamma\over f^{\rm (EDGES)}_a\right) .
\end{equation}
The energy fraction of the CMB in the EDGES frequency range is $f^{\rm (EDGES)}_\gamma$, and the $E_a \simeq (0.2$--$0.4)$ $\mu$eV is equivalent to the EDGES frequency range of 50 MHz to 100 MHz. The relation between energy $E$ and frequency $f$ is expressed as $E = 2\pi f \approx 4.14 \mu$eV$(f /1$ GHz).

The necessary excess effective number of neutrino species from the conversion probability of ALP to photon:
\begin{equation}
\Delta N_{\rm eff} \sim \left(f_a^{\rm (EDGES)}\over P_{a\; \to\; \gamma}\left(E_a\right)\right)\times 10^{-9} .
\end{equation}
We require at least the conversion probability of ALP to photon, $P_{a\; \to\; \gamma}(E_a)\approx 10^{-10}$ at the EDGES frequency range in order to eliminate the theoretical excess effective number of neutrino species, $\Delta N_{\rm eff}\sim 4$, from the cooling effect. This is evident by incorporating the $\Delta N_{\rm eff}$ cooling effect into $N_{\rm eff}$ in Eq.~(\ref{eqn4}) and resolving it via Eq.~(\ref{eqn5}). We found an extremely low ALP--photon conversion probability in comparison to prior research \cite{2018PhLB..783..301M}. The primordial intergalactic magnetic field is $\gtrsim$10$^{-17}$ G on Mpc scales, whereas primordial black hole's magnetic field strength is around $10^{-20}$--$10^{-15}$ G \cite{10.1093/mnras/stu357}. The exceedingly low mass of ALPs around $10^{-22}$ eV, along with a significantly reduced conversion rate, would contribute to the reduction of the potential excess effective number of neutrino species $\Delta N_{\rm eff}$ and explain the EDGES anomaly. Currently, $\Delta N_{\rm eff}$ must exceed 0.04; future enhancements in the limits on $\Delta N_{\rm eff}$ will serve as an additional assessment of our thermal equilibrium in Eq.~(\ref{eqn3}).

\section{Conclusions}

When primordial black holes and/or intergalactic magnetic fields are present, ALPs become metastable particles that eventually decay to photons. By decreasing $\Delta N_{\rm eff}$ and perhaps causing an asymmetry between the neutrino and antineutrino energy distributions, the heating and cooling effects can drastically change the predicted impact on primordial neutrinos. Our results require cosmological investigations on general new models of physics to be revised by integrating 21-cm line data with CMB polarization observations. We have thoroughly examined the effects on $\Delta N_{\rm eff}$ using the analysis provided in this contribution. Depending on the power spectrum of the primordial magnetic fields, this scenario might likewise leave distinctive traces on the 21-cm line fluctuation.

The viable parameter region of ALP--photon coupling has a strength that can be tested in future axion experiments like The International Axion Observatory (IAXO) and Axion Dark Matter eXperiment (ADMX), which can reach the sensitivity of $g_a =$ (a few)$\times 10^{-12}$ GeV$^{-1}$. Future CMB experiments like The Primordial Inflation Explorer (PIXIE) and Polarized Radiation Imaging and Spectroscopy Mission (PRISM) will greatly improve the constraints on energy release in the early universe, which may confirm or rule out our scenario on $\Delta N_{\rm eff}$. The Dark Ages Polarimeter PathfindER (DAPPER) and the Farside Array for Radio Science Investigations of the Dark Ages and Exoplanets (FARSIDE) will put the standard cosmological model to the test in an unexplored epoch, potentially revealing new physics involving novel interactions between baryons and ALPs.

\bibliography{crdas}
\end{document}